\documentclass[aip,reprint]{revtex4-1}

\usepackage{graphicx}% Include figure files
\usepackage{dcolumn}% Align table columns on decimal point
\usepackage{bm}% bold math
\usepackage{xcolor} 

\usepackage[colorlinks=true, linkcolor=blue, citecolor=blue]{hyperref}
\usepackage{epstopdf, epsfig}
\usepackage[export]{adjustbox}
\usepackage[amssymb]{SIunits}
\usepackage{amsfonts}
\usepackage{enumitem}
\usepackage{latexsym}
\usepackage{amssymb}
\usepackage[tight,footnotesize]{subfigure}
\usepackage{booktabs}
\usepackage{multirow}
\usepackage[utf8]{inputenc}
\usepackage{longtable}
\usepackage{array}
\usepackage{wasysym}
\usepackage{placeins}

\usepackage{comment}

\newcommand{\fig}{\mbox{Fig.~}}

\begin{document}

\title{Twente Mass and Heat Transfer Water Tunnel: Temperature controlled turbulent multiphase channel flow with heat and mass transfer}

\author{Biljana Gvozdi\'c}
\author{On-Yu Dung}
\author{Dennis P. M. van Gils}
\author{Gert-Wim H. Bruggert}
\affiliation{Physics of Fluids Group, J. M. Burgers Center for Fluid Dynamics and Max Planck Center Twente, Faculty of Science and Technology, University of Twente, P.O. Box 217, 7500 AE Enschede, The Netherlands}
\author{Elise Alm\'eras}
\affiliation{Laboratoire de G\'{e}nie Chimique, UMR 5503, CNRS-INP-UPS, 31106 Toulouse, France}
\author{Chao Sun}
\affiliation{Center for Combustion Energy, Key Laboratory for Thermal Science and Power Engineering of Ministry of Education, Department of Energy and Power Engineering, Tsinghua University, 100084 Beijing, China}
\affiliation{Physics of Fluids Group, J. M. Burgers Center for Fluid Dynamics and Max Planck Center Twente, Faculty of Science and Technology, University of Twente, P.O. Box 217, 7500 AE Enschede, The Netherlands}
\author{Detlef Lohse}
\affiliation{Physics of Fluids Group, J. M. Burgers Center for Fluid Dynamics and Max Planck Center Twente for Complex Fluid Dynamics, Faculty of Science and Technology, University of Twente, P.O. Box 217, 7500 AE Enschede, The Netherlands}
\affiliation{Max Planck Institute for Dynamics and Self-Organization, Am Fa{\ss}berg 17, 37077 G\"ottingen, Germany}
\author{Sander G. Huisman}
\email{s.g.huisman@utwente.nl}
\affiliation{Physics of Fluids Group, J. M. Burgers Center for Fluid Dynamics and Max Planck Center Twente, Faculty of Science and Technology, University of Twente, P.O. Box 217, 7500 AE Enschede, The Netherlands}
\date{\today}

\begin{abstract}
A new vertical water tunnel with global temperature control and the possibility for 	 bubble and local heat \& mass injection has been designed and constructed. The new facility offers the possibility to accurately study heat and mass transfer in turbulent multiphase flow (gas volume fraction up to $8\%$) with a Reynolds-number range from $1.5 \times 10^4$ to $3 \times 10^5$ in the case of water at room temperature. The tunnel is made of high-grade stainless steel permitting the use of salt solutions in excess of 15$\%$ mass fraction. The tunnel has a volume of \unit{300}{\liter}. The tunnel has three interchangeable measurement sections of $1$ m height but with different cross sections ($\unit{0.3 \times 0.04}{\meter ^2}$, $\unit{0.3 \times 0.06}{\meter ^2}$, $\unit{0.3 \times 0.08}{\meter ^2}$). The glass vertical measurement sections allow for optical access to the flow, enabling techniques such as laser Doppler anemometry, particle image velocimetry, particle tracking velocimetry, and laser-induced fluorescent imaging. Local sensors can be introduced from the top and can be traversed using a built-in traverse system, allowing for e.g. local temperature, hot-wire, or local phase measurements. Combined with simultaneous velocity measurements, the local heat flux in single phase and two phase turbulent flows can thus be studied quantitatvely and precisely.
\end{abstract}

\maketitle %\maketitle must follow title, authors, abstract and \pacs

%%%%%%%%%%%%%%%%%%%%%%%%%%%%%%%%%%%%%%%%%%%%%%%
\section{\label{sec:intro}Introduction}
%%%%%%%%%%%%%%%%%%%%%%%%%%%%%%%%%%%%%%%%%%%%%%%

Bubbles injected in turbulent flows are widely used in industrial processes to enhance mixing of the continuous phase and consequently the overall heat and mass transfer\cite{mudde2005,risso2018,deen2010}.
In practical applications two phase flows are preferred for example for highly exothermic processes where heat removal restricts the reactor's performance or for processes where the rate of mass transfer between a gas and a liquid is limited by the diffusion of a solute in the liquid\cite{darmana2006}. Another benefit of injection of bubbles is that no additional moving mechanical parts\cite{Jain2013} are needed for enhanced mixing which leads to relatively high reliability, and low maintenance and operating costs. Furthermore, bubble columns\cite{Besagni2018} are easier to use at higher temperatures and pressures when shaft sealing may be difficult\cite{deckwer1980mechanism}. 

The wide range of applications of bubbly flows has led to many studies on understanding the influence of bubble injection on heat and mass transfer in different configurations\cite{Deckwer1980,heijnen1984,Kulkarni2007}. While the focus of application-oriented studies is to formulate empirical correlations for the net heat and mass transfer coefficients, fundamental research focuses on measuring and characterising the local flow statistics which leads to physical insight behind the observed correlations. In particular, recent studies in turbulent bubbly flows have investigated a variety of aspects such as: (i) bubble size and velocity distributions \cite{kitagawa2013natural, colombet2015dynamics}, (ii) global heat and mass transport measurements\cite{tokuhiro1994natural,ayed2007hydrodynamics,colombet2015dynamics}, (iii) computational studies with ideal boundary conditions \cite{deen2013direct,dabiri2015heat}, (iv) liquid velocity and temperature profile\cite{kitagawa2013natural,gvozdic2018experimental,gvozdic2018experimental2},  (v) homogeneous and inhomogeneous bubble injection \cite{kitagawa2008heat,gvozdic2018experimental, gvozdic2018experimental2}, and (vi) natural and forced convection \cite{sekoguch1980forced, sato1981momentum, sato1981momentum2,kitagawa2008heat,kitagawa2009effects,dabiri2015heat}.

In systems of natural convection with bubble injection, mixing is provided by large scale circulations driven by density differences in the liquid and bubbles. 
Within such a system, studies have shown that the bubble size has a major effect on the overall heat transfer \cite{kitagawa2013natural}. 
However, there have been only a few studies on forced convective heat transfer in bubbly flows \cite{sekoguch1980forced,kitagawa2010experimental,dabiri2015heat}, where, in addition to buoyancy driven circulation, bubble wakes and their interplay provide an additional mixing mechanism. 
In systems of forced convection, experiments and numerical simulations have primarily focused on understanding the influence of bubble accumulation and deformability on the global mixing properties.
Experimental setups (e.g.~Ref.\cite{kitagawa2010experimental}) are generally limited by Reynolds number ($O(10^2)$). Industrial systems involving heat and mass transfer with forced convection, i.e. mean liquid velocity (for e.g. heat exchangers) reach much higher Reynolds numbers $O(10^3-10^5)$ and beyond.
Fundamental studies on the interaction between the turbulence in the carrier fluid, heat transfer and the dispersed bubbles at such high Reynolds numbers are currently lacking. The objective of our work is to fill this gap. In this paper we will describe a facility which is built to tackle various unresolved questions related to heat transfer in turbulent bubbly flow. 
The Twente Mass and Heat Transfer Water Tunnel will be used to: (i) quantitatively characterize the global heat transfer of a turbulent flow with and without gas bubble injection; (ii) correlate and understand the local heat flux with the local liquid velocity and temperature fluctuations in the bubbly turbulent flow; (iii) explore and understand the dependence of the heat transfer on the control parameters, such as the gas concentration, bubble size, and the Taylor--Reynolds number of the flow.

Another unexplored aspect of heat transport in bubbly flow is the influence of salt in the continuous phase which is highly relevant in certain industrial applications. 
For example, chlorate processes in chemical industry use brine (high molarity NaCl solution) as the continuous phase and the interaction of bubbles with turbulence and heat transfer in such systems is very different from bubbles in turbulence without salt. Dissolved salt greatly changes the interfacial properties of the bubbles thus leading to different forces between bubbles and thus different collision characteristics.
Even a small concentration of salt dissolved in a bubbly flow can thus significantly change the properties of the bubbles which may lead to drastic changes in the overall flow properties \cite{nguyen2012influence}. 
In order to study such flows and understand the influence of  salt (in particular NaCl) on the overall heat and mass transfer, we have built our setup specifically from corrosion-resistant stainless steel. 
In particular,  we aim to: (i) understand the effect of the salt concentration on the surface properties of bubbles and the resulting change of coalescence and breakup of them, and (ii): to quantify the resulting dependence of the global and local heat transport on salt concentration.

Apart from the possibility of heat transfer investigation, also mass transfer measurements can be performed in the new vertical tunnel. 
The study of the transport of microparticles in turbulent bubbly flow may shed light on finding the optimal parameters for the mixing of particles in a flow with for example catalytic particles, thereby ultimately maximising the overall efficiency of the chemical reactions in a chemical reactor. 
Thus, we aim to find the transport mechanism of a scalar field (e.g. the concentration field of the catalytic microparticles) in a turbulent bubbly flow. 

%%%%%%%%%%%%%%%%%%%%%%%%%%%%%%%%%%%%%%%%%%%%%%%
	\section{System description\label{sec:system_description}}
%%%%%%%%%%%%%%%%%%%%%%%%%%%%%%%%%%%%%%%%%%%%%%%

\begin{figure*}[t]
	\includegraphics[width=\textwidth]{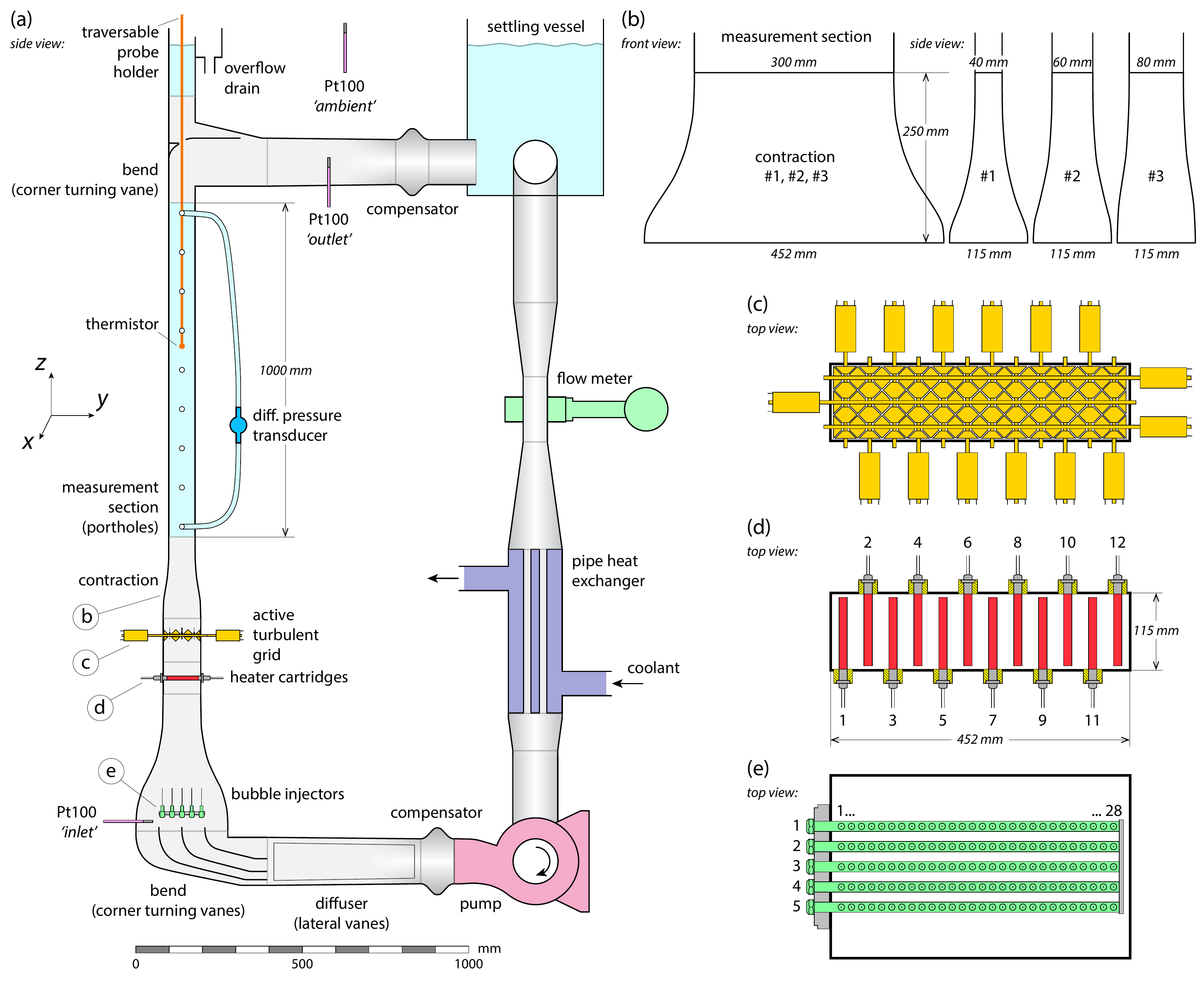}
	 \caption{\label{fig:setup}Schematic of the Twente Mass and Heat Transfer Water Tunnel  (TMHT). $x$, $y$, and $z$ are the spanwise, wall-normal, and streamwise directions, respectively, where each of the measurement sections has dimensions $W$ (width), $D$ (depth), and $L$ (length) in the $x$, $y$, and $z$ directions.
		(a) A side view cross-section of the tunnel showing the major components. 
		(b) The front and side profiles of the three exchangeable measurement sections and their respective contractions.
		(c) A top view of the active turbulent grid showing the fifteen independently rotating rods with agitator flaps attached to them, here all in their horizontal position for displaying purposes.
		(d) The twelve cylindrical heater cartridges in red and the thermal insulation in yellow.
		(e) Bubble injection provided by five rows of each twenty-eight exchangeable capillaries. The rows can be opened and closed independently from each other. 
		See Sec. \ref{sub:overview} for a walk-through description.}
\end{figure*}

\begin{figure}
	\includegraphics[width=0.9\columnwidth]{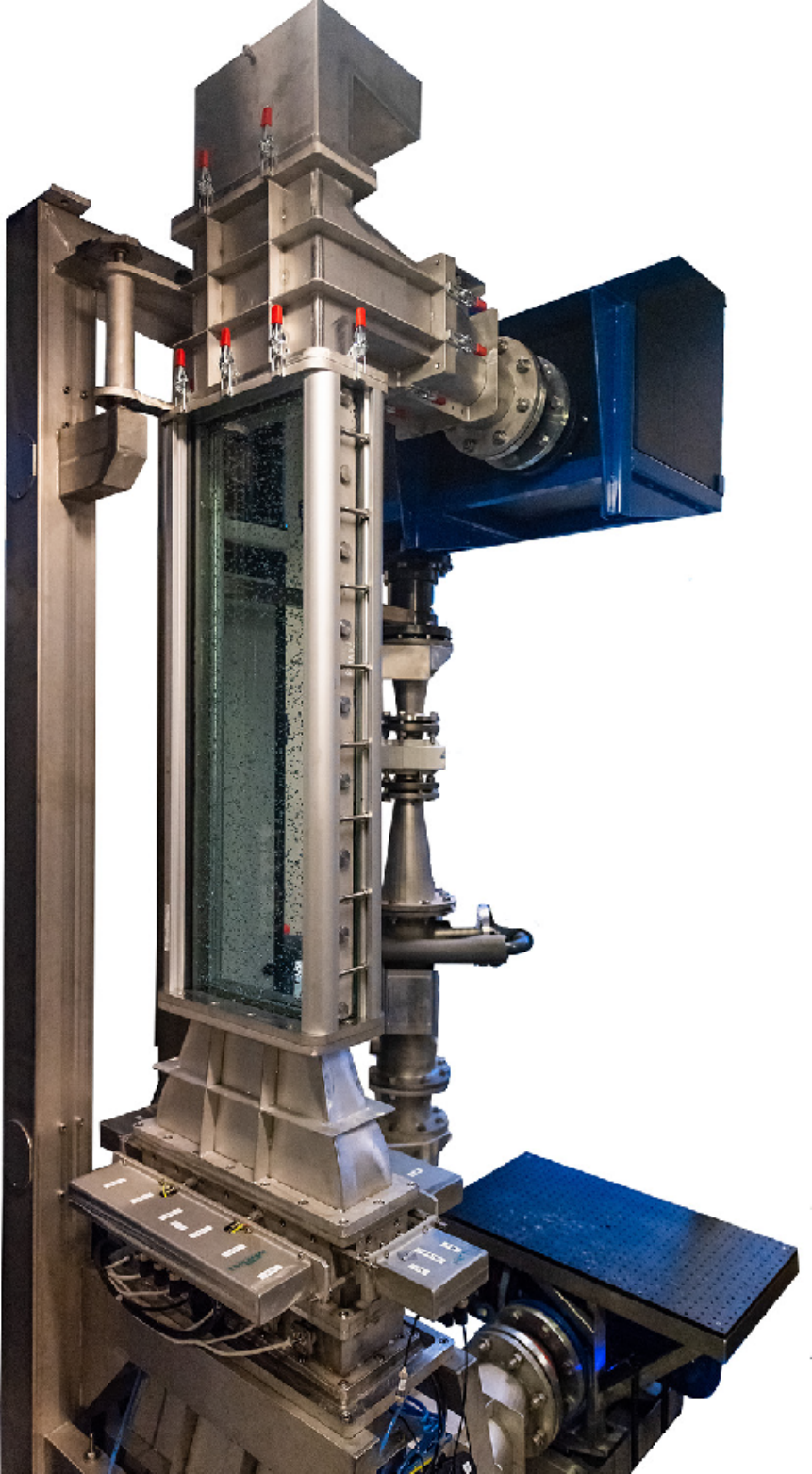}
	\caption{\label{fig:setupreal} The Twente Mass and Heat Transfer Water Tunnel (TMHT) in the lab. In the front (from bottom to top): the heaters, active grid and the measurement section. In the back (from top to bottom): settling vessel, flow meter, the heat exchanger and the pump.}
\end{figure}

The Twente Mass and Heat Transfer Water Tunnel (TMHT for short, see Figs.~\ref{fig:setup} and \ref{fig:setupreal}) is a recirculating vertical water tunnel for the study of heat and mass transfer in turbulent multiphase flows for both global and local quantities. 
We first list the main features:

\begin{itemize}
    \setlength{\itemsep}{0pt}
	\setlength{\parskip}{0pt}
	\setlength{\parsep}{0pt} 
    \item The tunnel has an internal volume of \unit{300}{\liter}.
    
    \item It is constructed out of marine-grade AISI316 stainless steel permitting the use of salt solutions in excess of $15\%$ mass fraction.
    
    \item An $\unit{80}{\meter ^3 \per \hour}$ capacity propeller pump drives the flow upwards in the measurement section from around $\unit{0.05}{\meter \per \second}$ up to $\unit{1}{\meter \per \second}$.
    
    \item Up to $\unit{10}{\kilo \watt}$ of heating power can be added into the flow by means of heater cartridges.
    
    \item A  $\unit{12.5}{\kilo \watt}$ chiller removes the added heat further downstream and allows for global temperature control of the measurement liquid within $\unit{20}{\milli \kelvin}$ long-term stability at statistically stationary conditions, see \fig \ref{fig:system_stability}b.
    
    \item There are three interchangeable measurement sections of each $\unit{1}{\meter}$ in length and of different cross sections, where the width $\times$ depth are either $\unit{0.3 \times 0.04}{\meter ^2}$, $\unit{0.3 \times 0.06}{\meter ^2}$, or $\unit{0.3 \times 0.08}{\meter ^2}$.
    
    \item Using water at $\unit{21}{\degree \Celsius}$ and a flow velocity range covering $\unit{0.05}{\meter \per \second}$ to $\unit{1}{\meter \per \second}$ in the measurement section results in a Reynolds number range of $\mathrm{Re} = 1.5 \times 10^4$ to $3 \times 10^5$, where $\mathrm{Re}=UW/\nu$ with $U$ the mean flow velocity, $W$ the width of the measurement section, and $\nu$ the kinematic viscosity of the liquid.
    
    \item Three of the four walls of each measurement section are made of glass (front, back, and one side) in order to gain optical access to the flow allowing techniques such as high-speed imaging, laser-Doppler anemometry, particle image velocimetry, particle tracking velocimetry and laser-induced fluorescent imaging. The fourth wall has several portholes where probes that measure local flow properties can be inserted.
        
    \item In the case of heat transfer studies, the tunnel is intended to operate at statistically stationary conditions with the tunnel liquid temperature around the ambient lab temperature to ensure minimal heat flux through the walls.
    
    \item Millimetric bubbles can be injected into the flow by 140 exchangeable capillaries. The gas volume fraction inside the measurement section can be measured electronically and can go up to $8\%$, depending on the flow conditions.
    
    \item An active turbulent grid consisting of agitator flaps that rapidly rotate inside the flow  agitates the turbulence inside the measurement section to achieve high turbulent intensity levels.
    
    \item A computer--controlled two--axis traversing frame allows for positioning sensors inside of the measurement section in order to obtain local quantities, such as the local temperature by using thermistors with a few mK resolution, or the local gas volume fraction by using the optical fiber probe technique \cite{cartellier1990optical}.
    
    \item The local heat flux inside the flow can be measured by simultaneously acquiring local velocity with laser-Doppler anemometry and local temperature with a thermistor.
\end{itemize}

In the next sections we will describe the TMHT facility in more detail.
 
%----------------------------------------------------
\subsection{Overview\label{sub:overview}}
%----------------------------------------------------

We now give a quick walk-through of all the major components of the tunnel, see \fig \ref{fig:setup} for a schematic picture and \fig \ref{fig:setupreal} for a photo. 
Starting from the pump, the flow is expanded from a circular cross-section to a larger rectangular area by the diffuser that has two laterally adjustable vanes inside to help steer the flow sideways. 
The flow then enters the bend and is guided by three corner--turning vanes to facilitate uniformity in the downstream flow velocity profile. 

A Pt100 temperature probe measures the bulk `\emph{inlet}' liquid temperature. 140 exchangeable capillaries (detailed in \fig \ref{fig:setup}e) can inject bubbles into the flow near the bottom of this vertical tunnel section. 
Two modular sections follow that can be interchanged with each other. 
These are the section containing the heater cartridges (detailed in \fig \ref{fig:setup}d) and the section with the active turbulent grid (detailed in \fig \ref{fig:setup}c). 
A contraction placed downstream of these sections prevents the flow separation; it contracts the flow over its width ($x$-direction) and depth ($y$-direction).

Three interchangeable measurement sections of different cross sections can be installed. Their width and depth are either $\unit{0.3 \times 0.04}{\meter ^2}$, $\unit{0.3 \times 0.06}{\meter ^2}$, or  $\unit{0.3 \times 0.08}{\meter ^2}$, each with a matching contraction of height $\unit{0.25}{\meter}$ (detailed in \fig \ref{fig:setup}b). 
All those measurement sections are 	$\unit{1}{\meter}$ tall and their front, back and one of the sides are made of glass. 
The remaining sidewall is of stainless steel and has equidistant portholes for inserting probes or injecting dye and a wet/wet differential pressure transducer is attached to two of these portholes to monitor the gas volume fraction. 
The top part of the tunnel directly above the measurement section is open and this can be used to insert a long probe holder, with e.g.~a thermistor attached to its end, down into the measurement section. 
The pole is attached to a computer--controlled two-axis traversing frame (not shown), allowing for automated positioning over the $x$ and $z$-directions. 
The $y$-direction can be traversed manually by a positioning stage. 
At the top of the measurement section the flow goes through a bend after which we find another Pt100 temperature probe to register the bulk `\emph{outlet}' liquid temperature. 
Another Pt100 probe on the outside of the tunnel monitors the ambient air temperature of the lab. 
The flow then enters the settling vessel where bubbles can rise up to the surface or where salt can be mixed into the liquid. 
Going downwards from the vessel the flow is sped up by a funnel to pass through an electromagnetic flow rate meter. 
Lastly, all heat that was added by the heater cartridges can be removed again by the pipe heat exchanger which is fed with coolant from a chiller, after which the flow enters the pump again.

%----------------------------------------------------
\subsection{Flow control and adjustment\label{sub:flow_control}}
%----------------------------------------------------

The flow in the tunnel is driven by a custom close-coupled axial flow propeller pump (Herborner Pumpenfabrik, Uniblock P 125-201/0074) fully cast from marine-grade AISI316 stainless steel to allow the use of saline solution. 
The capacity is 80 m$^3$/h at a maximum power of \unit{0.75}{\kilo\watt} and \unit{1500}{rpm}. 
The three-phase electric motor of the pump is controlled by a digital frequency inverter (Herborner Pumpenfabrik, PED).

Directly after the pump is a rubber compensator to accommodate for thermal expansion of the tunnel. 
A diffuser with a circular inlet with a diameter of \unit{136}{\milli\meter} expands the flow to a rectangular outlet of \unit{452 \times 146}{\milli\meter^2}. 
Inside the diffuser are two adjustable lateral vanes to help spread the flow sideways. 
The next bend directs the flow upwards. 
Inside the bend are three adjustable corner turning vanes that guide the flow and help increase flow velocity uniformity after the bend.

An electromagnetic flow meter (ABB, ProcessMaster FEP311) with a bore \unit{65}{\milli\meter} measures the volumetric flow rate over its nominal range of \unit{2.4}{\meter^3/hour} to \unit{120}{\meter^3/hour}. According to the manual the measurement accuracy is better than $1\%$ if the flow velocity through the meter is larger than 0.2 m/s, which is the reason for choosing this specific bore diameter. 
The flow is sped up by a funnel with an inclination angle of  \unit{8}{\degree} before entering the meter.

The flow rate reported by the flow meter is continuously read out by an Arduino M0 Pro microcontroller board over a $4$--$\unit{20}{\milli \ampere}$ current loop. 
A tuned proportional-integral controller programmed into the Arduino feeds back to the propeller rotation rate of the tunnel pump over another $4$--$\unit{20}{\milli \ampere}$ current loop to ensure a stable flow rate and velocity over time within $1\%$ of the setpoint, see \fig\ref{fig:system_stability}e and \fig\ref{fig:system_stability}f.

%----------------------------------------------------
\subsection{Temperature control and heat injection\label{sub:temperature_control}}
%----------------------------------------------------

\begin{figure*}[t]
	\includegraphics[width=\textwidth]{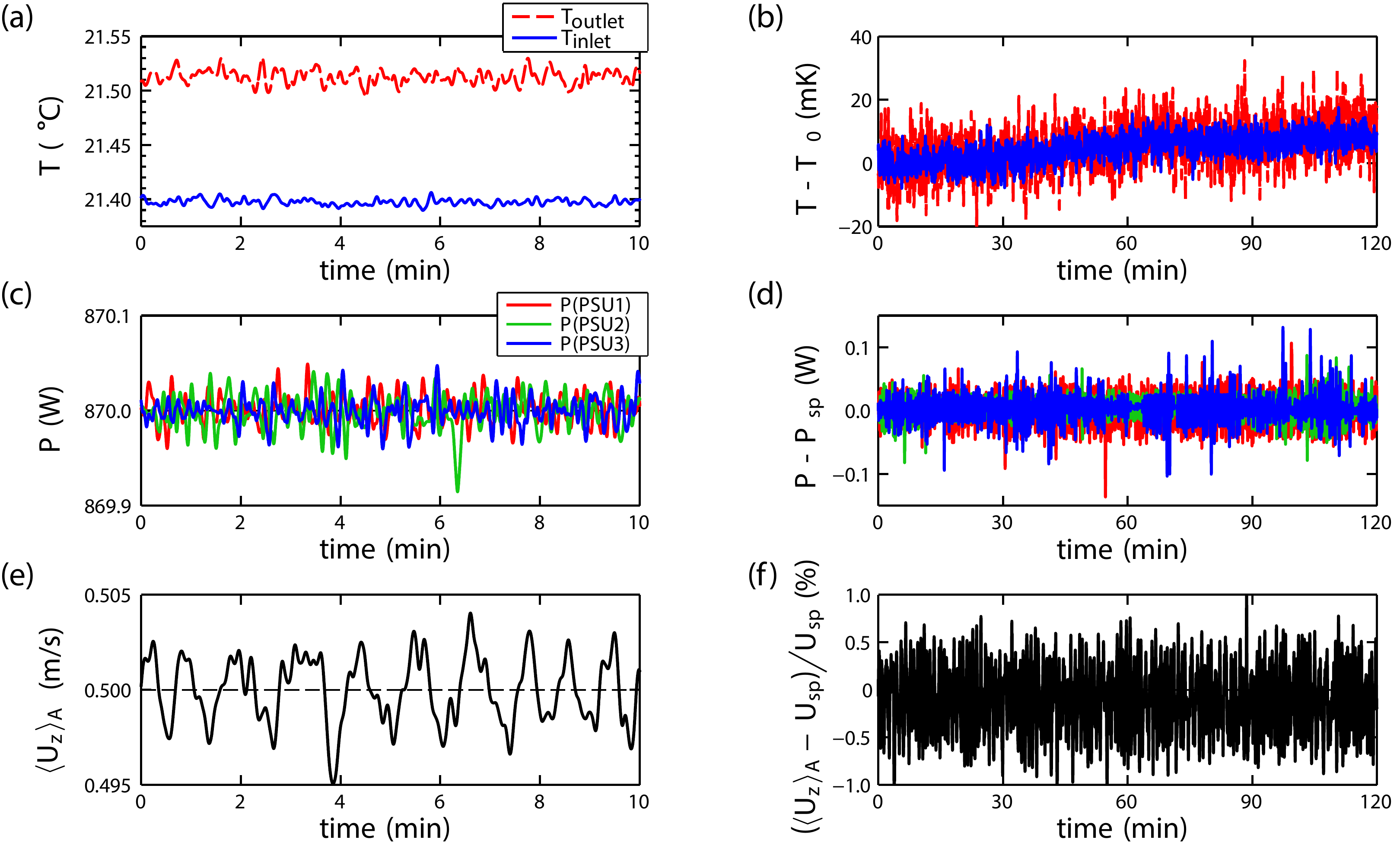}
		\caption{\label{fig:system_stability} Typical timeseries of the main global tunnel quantities after having settled to statistically stationary conditions, demonstrating the short-term fluctuations in (a), (c) and (e) and the long-term stability in (b), (d) and (f). This specific case was at a set tunnel flow velocity of $\mathrm{U}_{sp} = 0.5$ m/s using measurement section $\#$1 (subscript sp stand for setpoint), a chiller setpoint of \unit{19}{\degree\Celsius} and a set heating power of $\unit{3 \times 870}{\watt} = \unit{2610}{\watt}$ divided over heaters 1 to 6. Heaters 1 and 2 were connected in series to power supply PSU1, likewise 3 and 4 to PSU2 and 5 and 6 to PSU3. This caused the heaters to operate at an internal temperature of near \unit{90}{\degree\Celsius}.
		(a) Temperatures $\mathrm{T}$ of the tunnel inlet (solid blue) and outlet (dashed red).
		(b) The same data as the left panel, but with the starting temperature $\mathrm{T}_0$ subtracted. The temperature drift is below 20 mK over 120 minutes.
		(c) Heating power $\mathrm{P}$ provided by power supplies PSU1 (red), PSU2 (green) and PSU3 (blue).
		(d) The same data as the left panel, but with the setpoint of the power $\mathrm{P}_{sp}$ subtracted. The power is stable within 0.15 W regardless of the setpoint.
		(e) Instantaneous streamwise flow velocity inside the measurement section $\langle\mathrm{U}_z\rangle_A$ averaged over its cross-sectional area $A$ (measured using the flow meter). The dashed line indicates the setpoint.
		(f) The same data as the left panel, but renormalized with the flow velocity setpoint $\mathrm{U}_{sp}$. The flow velocity is stable within 1$\%$ of the setpoint without drift over 120 minutes.}
\end{figure*}

Heat can be injected into the flow by twelve cylindrical heater cartridges (Watlow, Firerod J5F-15004) placed \unit{300}{\milli \meter} below the measurement section and sticking perpendicular to the flow direction through the front and back tunnel walls, see \fig \ref{fig:setup}d. 
Each heater cartridge has a diameter of \unit{12.7}{\milli \meter} and a heated zone of \unit{110}{\milli \meter} long as measured from its tip inside the tunnel, indicated by red. 
The end of the heated zone matches the start of the interior tunnel wall. 
Each heater is thermally insulated from the wall by a Teflon (PTFE) feed-through, indicated by yellow. 
A J-type thermocouple with an accuracy of \unit{\pm2.2}{\kelvin}, for the absolute temperature, is embedded inside of each heater. 
Because the location of the thermocouple is not guaranteed to be perfectly centred and a strong temperature gradient is to be expected in the interior of the heater cartridge when powered and placed in a flow, the thermocouple temperature readings are only to be used as a rough indication of the heater temperature. Hence, they are solely used for over-temperature protection. 
The heaters are rated for temperatures in excess of \unit{250}{\degree \Celsius}, but will be limited by the over-temperature protection to a maximum of \unit{95}{\degree \Celsius} to prevent local boiling.

The heaters are operated by controlling the supply power. 
Each heater can provide \unit{1160}{\watt} of heat at a maximum of \unit{120}{\volt}. 
They are powered by three independently programmable DC power supplies (Keysight, N8741A), each providing up to \unit{3.3}{\kilo \watt} at \unit{300}{\volt} and \unit{11}{\ampere}. 
Either a single heater is connected to a single power supply, or multiple heaters are connected in series to a single power supply, and any combination of this depending on the needs of the experiment. 
The programming accuracy of the power supplies is \unit{150}{\milli \volt} and \unit{22}{\milli \ampere}, and the measurement accuracy is \unit{300}{\milli \volt} and \unit{33}{\milli \ampere}. 
A tuned proportional-integral controller running on a computer regulates the power supply output voltage in order to tune the power output, resulting in an accuracy of the set power of \unit{\pm 0.3}{\watt} with a settling time of below \unit{10}{\second}. The power is stable within \unit{0.15}{\watt} regardless of the setpoint, see \fig\ref{fig:system_stability}c and \fig\ref{fig:system_stability}d.

The bulk temperature of the liquid inside the tunnel is measured at two locations using two Pt100 temperature probes with a 1/10 DIN accuracy corresponding to \unit{\pm 30}{\milli\kelvin} (Pico Technology, PT-104 data logger with SE012 probes). 
One probe labelled `\emph{inlet}' is upstream of the heaters at the location of the bubble injectors and protrudes \unit{100}{\milli \meter} into the flow, see \fig \ref{fig:setup}a. 
The other probe labelled `\emph{outlet}' is downstream of the measurement section and in front of the settling vessel. 
A third Pt100 probe labelled `\emph{ambient}' measures the ambient lab air temperature.

Cooling of the tunnel liquid is taken care of by a pipe heat exchanger, manufactured on request by Geurts International, located just in front of the tunnel pump intake. Twenty-two marine-grade AISI316 stainless steel pipes of \unit{500}{\milli \meter} long and with diameters of \unit{22}{\milli \meter} through which the tunnel liquid is flowing, are embedded in an outer jacket through which cooling water is flowing. The pipe heat exchanger is designed to remove up to \unit{10}{\kilo \watt} of heat. The temperature of the cooling water is regulated by a \unit{12.5}{\kilo \watt} capacity air-cooled recirculating chiller (Thermo Scientific, ThermoFlex 15000 P5) with a listed temperature stability of \unit{\pm 0.1}{\kelvin}.

We will now show a typical example of settling times and temperature stability. Starting with a quiescent water-filled tunnel at room temperature, it takes up to 2.5 hours to settle to a statistically stationary state, from the moment of setting the flow rate to \unit{21.6}{\meter ^3 \per \hour} (equivalent to a velocity of \unit{0.5}{\meter \per \second} in the measurement section $\#$1), the global heat input to \unit{2.6}{\kilo\watt} and the setpoint of the chiller at \unit{19}{\degree \Celsius}. After this settling time, we report a long-term stability of the tunnel liquid temperature of less than\unit{\pm 10}{\milli \kelvin} for a duration of over 2 hours, see \fig\ref{fig:system_stability}.

%----------------------------------------------------
\subsection{Local temperature measurements\label{sub:local_measurements}}
%----------------------------------------------------

High precision and local temperature measurements can be made with thermistors. We use thermistors from TE Connectivity (Measurement Specialties G22K7MCD419) with a glass bead of diameter \unit{0.38}{\milli \meter}  and a listed response time of \unit{30}{\milli \second} in liquids. All thermistors are calibrated against the `\emph{inlet}' Pt100 probe with a 1/10 DIN accuracy, corresponding to \unit{\pm 30}{\milli \kelvin} (Pico Technology, PT-104 data logger with SE012 probe) inside of a large copper body placed in a temperature bath with a \unit{5}{\milli \kelvin} temperature stability (PolyScience, PD15R-30), resulting in a calibration accuracy of \unit{\pm 30}{\milli \kelvin} and a \unit{5}{\milli \kelvin} resolution.

% at an acquisition rate of up to 1000 Hz

Temperature time series can be acquired by using a dual-phase lock-in amplifier (Stanford Research, SR830) in combination with a balanced Wheatstone bridge of which one arm is a single thermistor. 
A digital acquisition card (National Instruments, PCI-6221) logs the amplifier's output voltage to a file on the computer. Multiple thermistors can be logged during a measurement by a digital multimeter (Keysight, 34972A) installed with a reed-relay multiplexer board (Keysight, 34902A) that will scan sequentially over all the thermistor resistance values with a listed scan rate of up to 250 channels per second.

%----------------------------------------------------
\subsection{Traverse\label{sub:traverse}}
%----------------------------------------------------

Accurately positioning thermistors, optical fibers or other probes inside of the measurement section and automatically scanning over multiple grid points is made possible by a custom-built two-axis traversing frame located above the tunnel. 
The traverse consists of two linear actuators (Parker, HMRB15SBD0-1800 and HMRB11SBD0-1000), powered by two servomotors (Parker, SMHA60451 and SMH60301) and digitally operated by two controllers (Parker, Compax3). 
The traverse can travel over the full width ($x$-direction) and full height ($z$-direction) of the measurement section. 
A probe holder is attached to the traverse carriage and descents into the measurement section from an opening in the top of the tunnel. 
The carriage can be positioned with an accuracy of \unit{\pm 0.1}{\milli \meter}. 
The third axis (depth, $y$-direction) can be traversed over manually by using the micro-stage that is in between the traverse carriage and the traverse pole.

%----------------------------------------------------
\subsection{Bubble injection and gas volume fraction\label{sub:bubble_injection}}
%----------------------------------------------------

Bubbles can be injected into the flow by the pressurised air lines that are fed through the wall of the tunnel located underneath the heater cartridges. 
The lines are pressurised at 8 bars and connected to five rows of twenty-eight Luer lock fittings, see \fig \ref{fig:setup}a and \fig \ref{fig:setup}e. 
The standardised Luer lock allows for a wide variety of needles/capillaries to be installed and easily replaced (e.g. Nordson precision stainless steel dispensing tips), ranging from inner diameters of \unit{0.2}{\milli \meter} to \unit{2}{\milli \meter}. 
The resulting size of the bubbles detaching from the capillaries may be influenced by the local velocity and shear of the surrounding liquid, surfactants, air flow rate, and the geometry of the tip of the capillaries. 
Each of the five rows can be opened or closed individually by computer controlled piston valves. A digital mass flow controller (Bronkhorst, F-111AC-50K-AAD-22-V) drives air through all five rows at up to 44 litres per minute. %In the future, a larger capacity mass flow controller can be installed parallel to the current one. Also, instead of needles or capillaries, porous plates or cylinders can be installed to provide micro-meter sized bubble injection.

\begin{figure*}
	\includegraphics[width=0.88\textwidth]{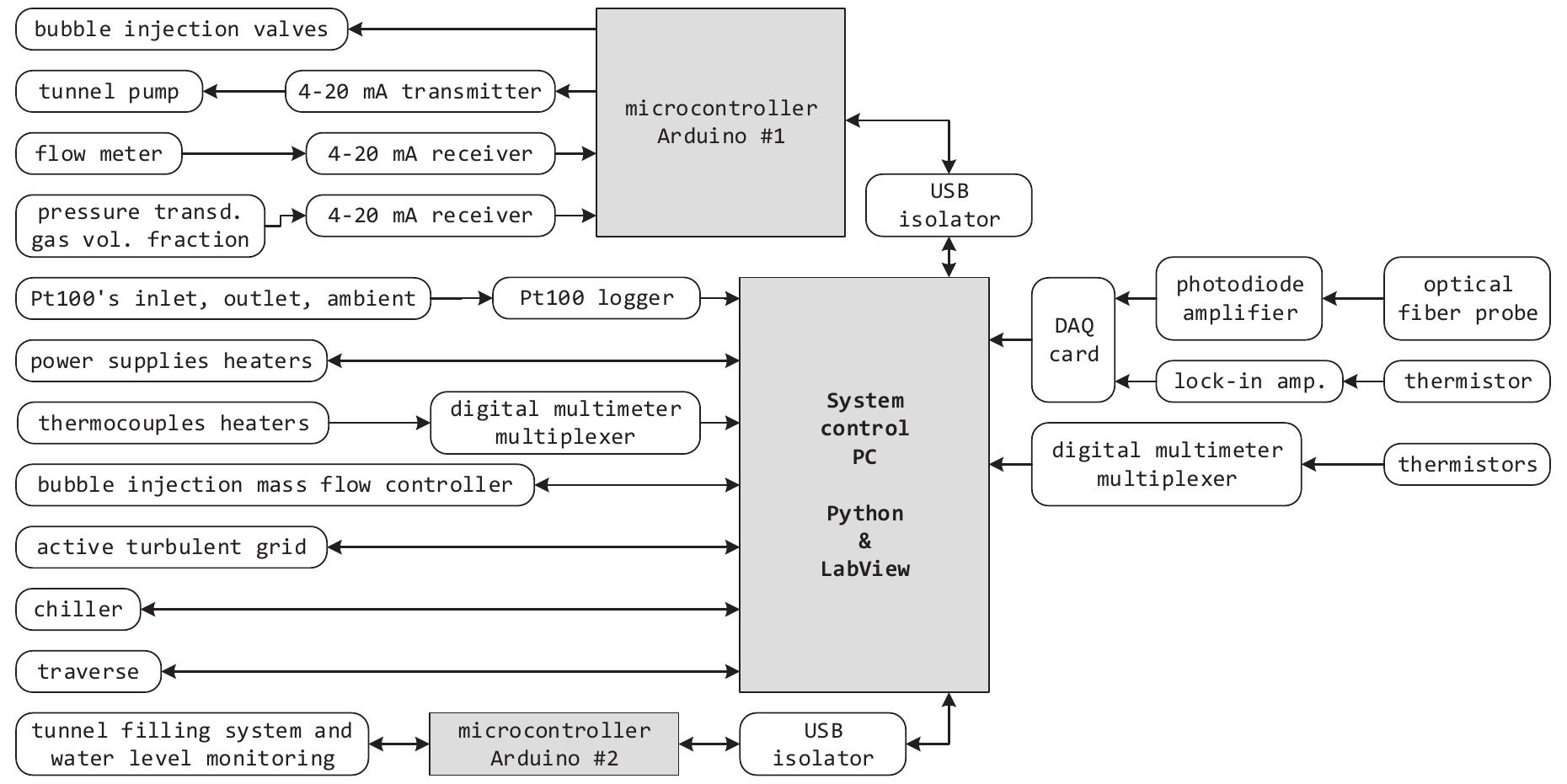}
	\caption{\label{fig:system_control_diagram} System control diagram. The tunnel filling system and water level monitoring running on Arduino $\#$2 are not discussed in this work and are mentioned for completeness.}
\end{figure*}

The global gas volume fraction over the full height of the measurement section is monitored by a wet/wet differential pressure transducer (Omega, PXM409170HDWUI). 
The pressure transducer is connected to two portholes, with a distance of $\Delta L = \unit{0.96}{\meter}$ in height apart, at the side of the measurement section. 
Each of these portholes has just a narrow channel of \unit{1.0}{\milli \meter} in diameter that connects to the tunnel's internal volume. 
This ensures that no air will enter the lines from the portholes towards the pressure transducer and that they are always filled with liquid. 
The resulting difference in static pressure between these lines and the liquid column inside the measurement section translates into the relation  $\alpha=\Delta P / (g\rho_l \Delta L$), where $\alpha$ is the global gas volume fraction over the measurement section, $\Delta P$ is the measured pressure difference, $g$ the local acceleration of gravity, $\rho_l$ the density of the liquid and $\Delta L$ the distance in height between the portholes. 
We report a constant measurement accuracy in $\alpha$ of $\pm 0.2\%$, given already in units of the gas volume fraction percentage. Injecting \unit{36}{\litre \per \minute} of air through the bubble injection capillaries, we can achieve a maximum of $\alpha = 7.8 \pm 0.2 \%$ when the tunnel pump is switched off, and e.g. $\alpha = 5.3 \pm 0.2 \%$ at a flow velocity of 0.5 m/s inside measurement section \#1.

The local gas volume fraction, bubble size and bubble velocity statistics can be obtained by the optical fiber probe technique. 
For details on this experimental technique and signal processing we refer to  Cartellier \cite{cartellier1990optical} and van Gils \emph{et al.} \cite{van2013importance}.

%----------------------------------------------------
\subsection{Active turbulent grid\label{sub:active_turbulent_grid}}
%----------------------------------------------------

The active turbulent grid consists of fifteen independently-rotating rods with agitator flaps attached to them that stir up the liquid flowing past, see \fig \ref{fig:setup}a and \fig \ref{fig:setup}c. 
It resembles the grid described in paragraph $\S$3.3.1 of Poorte and Biesheuvel\cite{poorte2002grid}, albeit with modern DC-motors (Maxon, DCX32L GB KL 24V) and controllers (Beckhoff, EL7332) and with only 15 rods instead of 24 because of the smaller cross-sectional area of the tunnel. 
There are several forcing protocols that can be used to drive the rods independently at varying rotation speeds and direction for varying periods of time. 
The study over the different protocols by Poorte and Biesheuvel\cite{poorte2002grid} revealed that the \emph{double-random asynchronous mode} protocol is the most favourable, as this protocol does not lead to periodicities in the turbulent power spectrum directly downstream of the grid. 
Hence, this protocol (P0238) is the standard we use in our studies. 
In short, it varies the rotation speed of a single rod and the duration of that rotation by randomly choosing values from the intervals $[-\Omega_m, \Omega_m]$ and [50, 90] ms, respectively. 
A control parameter called `\emph{grid speed factor}' $GSF$ will rescale the value of $\Omega_m$, such that the amount of agitation can be tailored. 
For example, the nominal value of $\Omega_m$ at a grid speed factor of 0.5 corresponds to \unit{18}{\hertz}.

%----------------------------------------------------
\subsection{System control\label{sub:system_control}}
%----------------------------------------------------

The system control of the TMHT facility is controlled by two programmable multifunction microcontroller boards in conjunction with a PC, see \fig \ref{fig:system_control_diagram} for a simplified diagram. 
The boards used are two Arduino M0 Pro which are powered by a SAMD21 microcontroller unit from Atmel, featuring a 32-bit ARM Cortex M0 core. 

One major source of electrical noise is carried by the lab facility electrical ground. 
Inductive spikes induced by e.g.~motors or switching relays can travel over the ground leads and can interfere with the sensitive microcontrollers. 
To prevent ground loops, ground noise, or inductive spikes from reaching the sensitive microcontrollers, one has to float the microcontroller board with respect to the lab facility ground potential. 
We do this by powering the microcontrollers using a noise-suppressing power supply (Traco Power, TCL 120-112) whose output is left floating with respect to ground. 
That means that ground-referenced peripheral devices cannot be connected to the Arduino directly, lest we loose the floating ground reference.
Hence, all USB connections from the microcontrollers to the PC are galvanically isolated by USB isolators (Olimex, USB-ISO). All peripheral digital sensors and actuators are connected to the microcontrollers with opto-couplers in between.

All analog signals from and to the microcontrollers (i.e. set pump speed, flow meter and read pressure transducer) transmit over 4--20 mA current loops, which are inherently insensitive to electrical noise in contrast to using voltage as a data carrier. 
The 4--20 mA current transmitter and receivers used are, respectively MIKROE-1296 T Click and MIKROE-1387 R Click from MikroElektronika. 
The remaining peripheral devices (i.e. the Pt100 data logger, programmable power supplies, digital multimeter multiplexers, mass flow controller, active turbulent grid, chiller and traverse controllers) are directly connected to the PC.

The main control program is running in Python 3.6 on the PC and handles all device input/output communication including the microcontrollers, with the exception of the active turbulent grid which is running on LabView from National Instruments. The main program has a graphical user interface to provide control, monitoring, and data logging of the TMHT facility, whose global quantities are logged at a rate of \unit{10}{\hertz}. The Python libraries that have been written to provide multithreaded communication with the specific laboratory devices including the Arduino microcontrollers are made available under the MIT open-source license and can be found at the GitHub repository \cite{github_vangils}.

%%%%%%%%%%%%%%%%%%%%%%%%%%%%%%%%%%%%%%%%%%%%%%%
\section{\label{sec:results}Examples of flow and temperature measurements}%
%%%%%%%%%%%%%%%%%%%%%%%%%%%%%%%%%%%%%%%%%%%%%%%

%----------------------------------------------------
\subsection{Velocity profile measurements}
%----------------------------------------------------
\begin{figure*}
	\centering
	\includegraphics[scale=0.4]{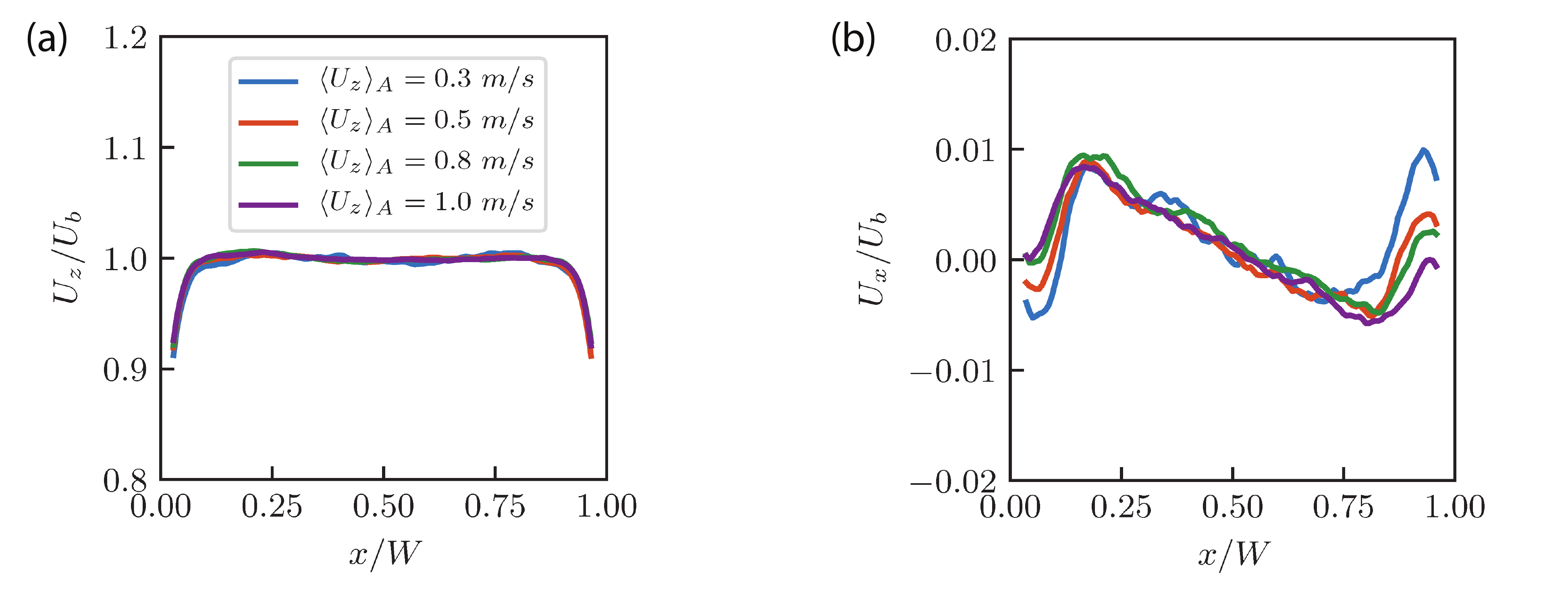}
	\caption{\label{fig:uz_mean}  Normalised (a) streamwise $U_z$ and (b) spanwise $U_x$ velocity profile along the width of the setup at half-height obtained from PIV measurements for different mean flow rates  $\langle U_z \rangle _A$ and constant speed of the active grid (the active grid is running at $ GSF=0.5$). $U_b$ is the streamwise velocity in the bulk, calculated by averaging the measured velocity over the width $x/W = 0.25$ to $x/W = 0.75$.}
\end{figure*}

\begin{figure*}
 	\includegraphics[scale=0.4]{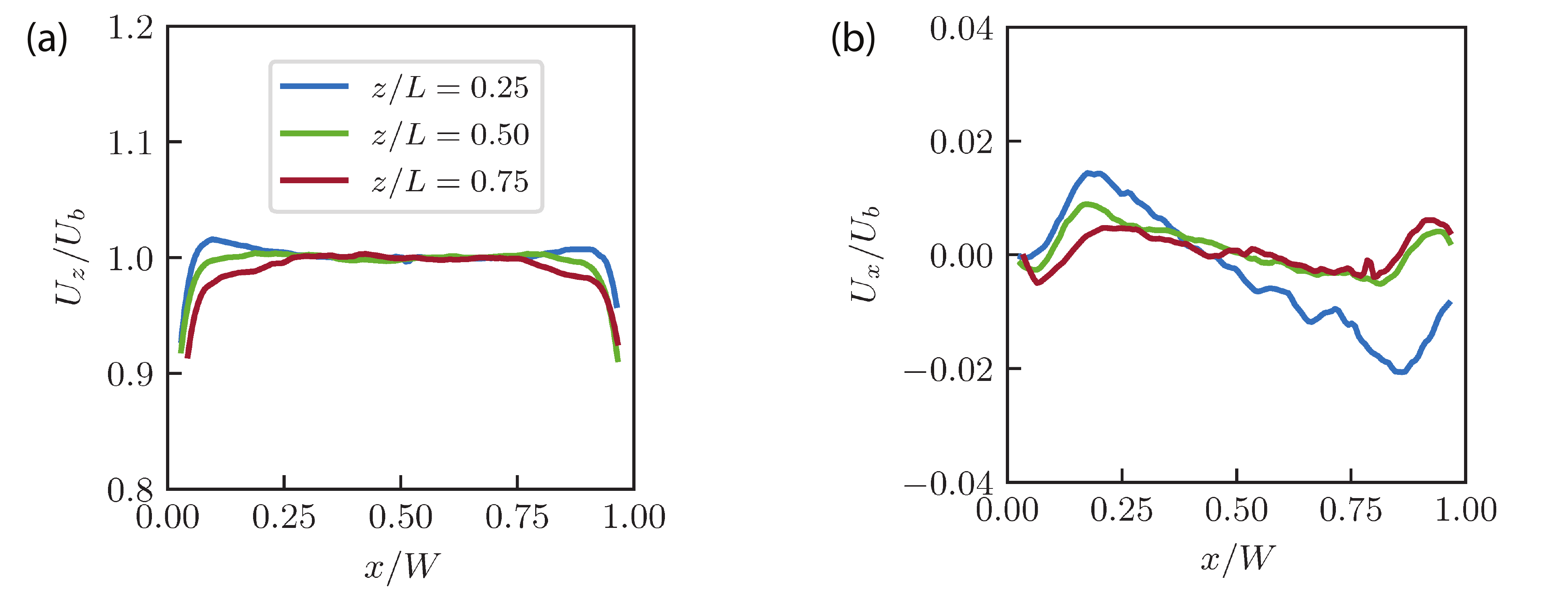}
 	\caption{\label{fig:uz_H} Normalised (a) streamwise $U_z$ and (b) spanwise $U_x$ velocity profile along the width of the setup obtained from PIV measurements at different heights for mean flow rate of $\langle U_z \rangle _A = \unit{0.5}{\meter \per \second}$ and constant speed of the active grid  $GSF =0.5$.  $U_b$ is the streamwise velocity in the bulk at each height, calculated by averaging the measured velocity over the width $x/W = 0.25$ to $x/W = 0.75$.}
\end{figure*}

\begin{figure*}
	\centering
	\includegraphics[scale=0.4]{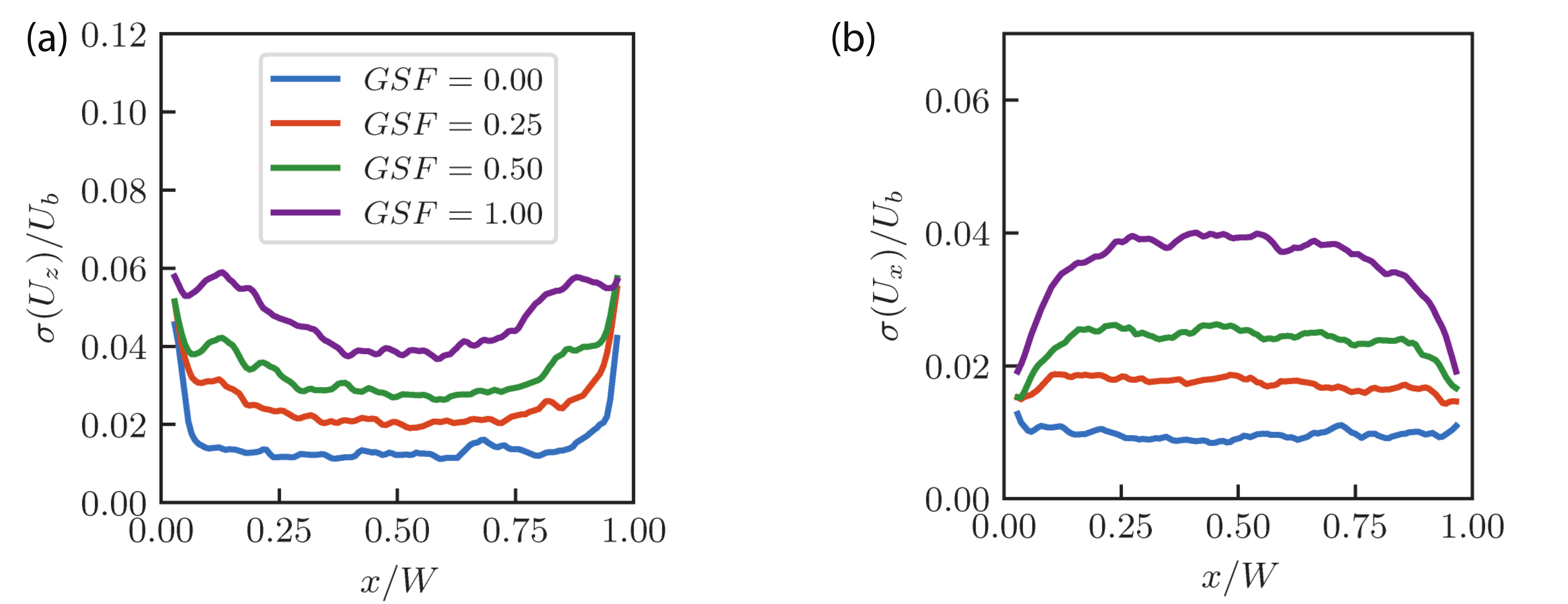}
	\caption{\label{fig:uz_std}  Turbulence intensity in (a) streamwise  and (b) spanwise direction along the width of the measurement section at half-height obtained from PIV measurements for $\langle U_z \rangle _A = \unit{0.5}{\meter \per \second}$ and varying speed of rotation of the active grid (varying the value of the grid speed factor ($GSF$)). $GSF=0$  corresponds to the measurement with the active grid switched off, while for $GSF=1$ rods are rotating at a maximum speed. $\sigma(U_z)$ is the standard deviation of the streamwise velocity, while $\sigma(U_x)$ is the standard deviation of the spanwise velocity.}
\end{figure*}

\begin{figure}
	\centering
	\includegraphics[scale=0.4]{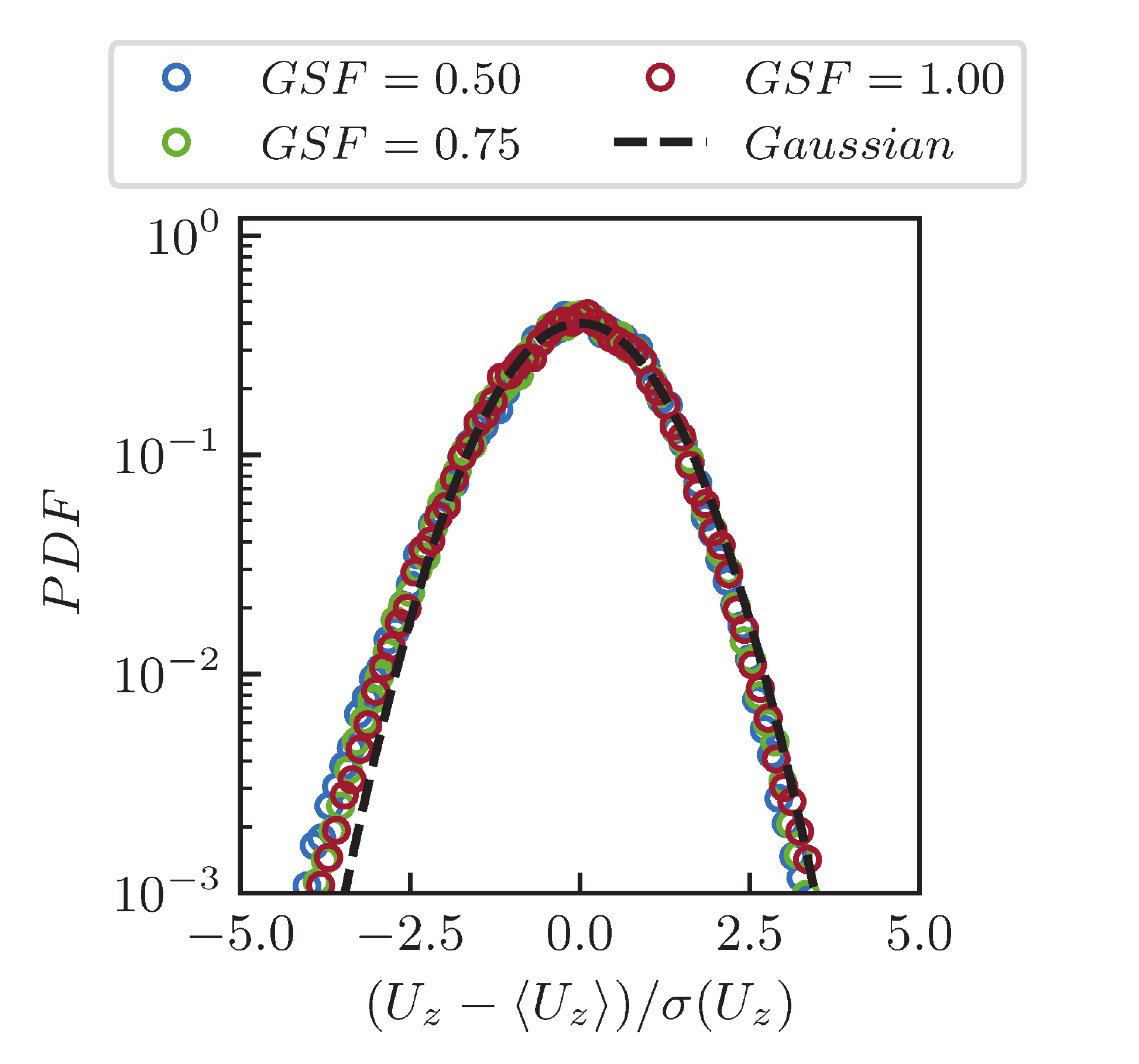}
	\caption{\label{fig:uz_pdf} Probability density function of the streamwise velocity for $\langle U_z \rangle _A = \unit{0.5}{\meter \per \second}$ and different values of the grid speed factor, obtained at $x/W = 0.5$, $y/D = 0.5$, $z/L = 0.5$ using LDA. Gaussian with zero mean and unit variance is added for reference.}
\end{figure}

\begin{figure}
	\centering
	\includegraphics[scale=0.4]{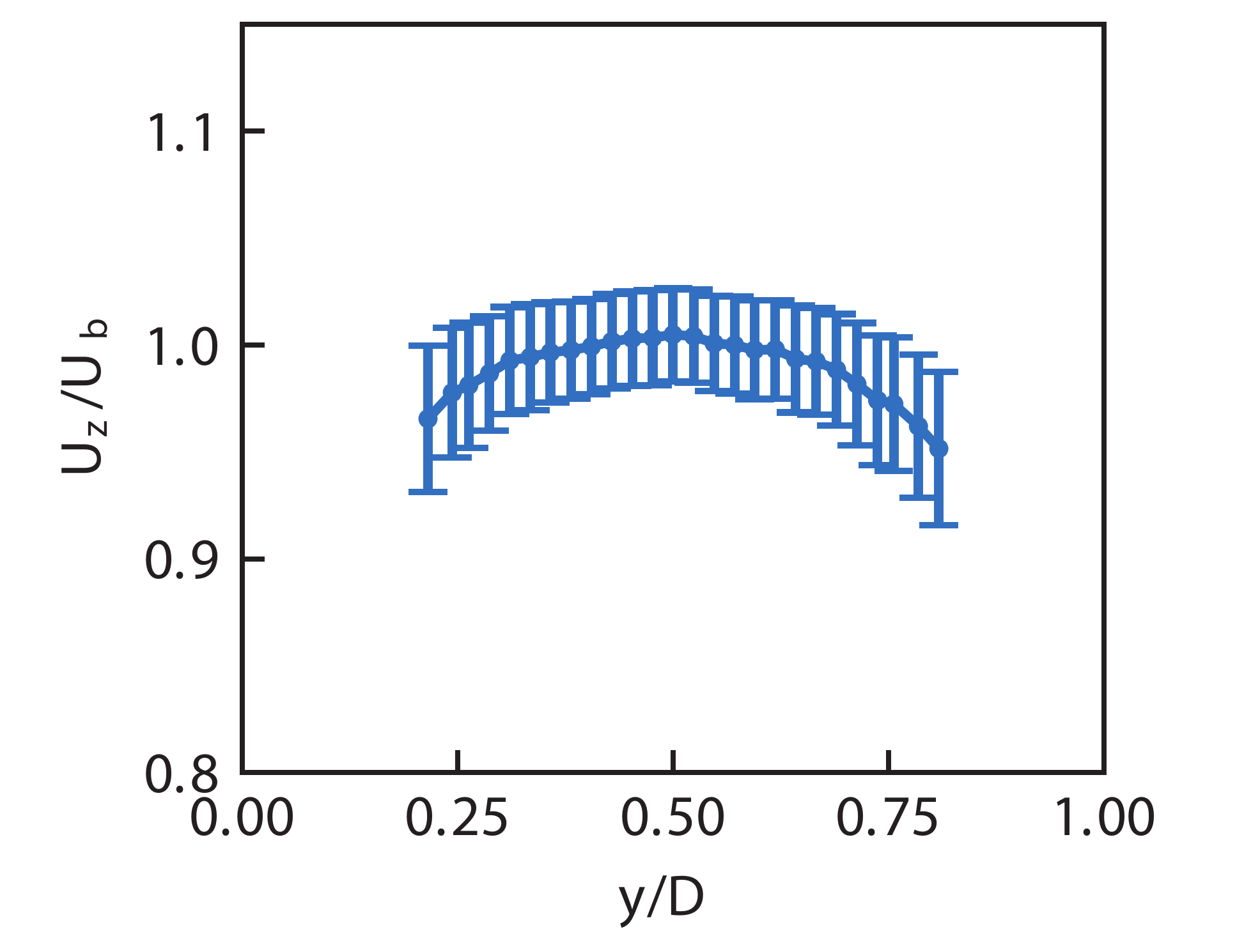}
	\caption{\label{fig:uz_width} Scan of the vertical velocity along the depth of the setup at half-height performed using LDA for $\langle U_z \rangle _A = \unit{0.5}{\meter \per \second}$ and $GSF = 0.5$. The bars corresponds to $\pm 1$ standard deviation of the velocity.}
\end{figure}

As a first step in demonstrating robustness and consistency of the setup, we measure the liquid velocity using two experimental techniques: laser Doppler anemometry (LDA) in backscatter mode and particle image velocimetry (PIV) in measurement section $\#$1 (width $\times$ depth: $0.3 \times \unit{0.04}{\meter ^2}$). 
For LDA measurements  the flow is seeded with polyamid seeding particles (diameter $\unit{5}{ \micro \meter}$, density $\unit{1050}{\kilo \gram \per \meter ^3} $). 
The LDA system used consists of DopplerPower DPSS (diode--pumped solid--state) laser and a Dantec burst spectrum analyser (BSA). 
For PIV we seed the flow with fluorescent tracer particles (diameter $ \unit{50}{ \micro \meter}$).
We  use a high speed laser (Litron LDY-303HE)  with a cylindrical lens to generate a sheet of light passing through the centre of the glass side-wall of the measurement section ($y/D = 0.5$).
A double-frame camera (PCO Imager sCMOS) at 20 fps was used to obtain the velocity field.
All the velocity profiles obtained using PIV presented in this section are calculated by averaging the velocity over several heights, covering $\pm \unit{5}{\centi \meter}$ around the noted height.

In figure \ref{fig:uz_mean} we plot the normalised mean profile of the streamwise and spanwise velocity obtained from PIV measurements for different mean flow rates, namely $\unit{0.3}{\meter \per \second}$, $\unit{0.5}{\meter \per \second}$, $\unit{0.8}{\meter \per \second}$, and $\unit{1.0}{\meter \per \second}$, for a constant active grid speed factor of 0.5.
We find that the variation of the mean horizontal and vertical velocity is negligible in the bulk of the setup, namely not more than $\pm 1.5\%$ from the mean streamwise velocity. 
In figure \ref{fig:uz_mean}b we see that the spanwise velocity has a positive and negative peak, which is a signature of the contraction section placed downstream of the measurement section. 

Next, we look at the velocity profiles for $\langle U_z \rangle _A =  \unit{0.5}{\meter \per \second}$ and $GSF = 0.5$ at different heights, namely at $z/L = 0.25$, $z/L = 0.5$ and $z/L = 0.75$. 
In figure \ref{fig:uz_H}a we see that the streamwise velocity remains constant (within $\pm 2\%$) along the width at different heights. 
By scanning the velocity at different heights we observe that the influence of the contraction on the spanwise velocity reduces as the height increases, although its amplitude is only of order of $1\%$ of the streamwise component (see Figure \ref{fig:uz_H}b).

In figure \ref{fig:uz_std} we plot the turbulence intensity measured using PIV at half-height for $\langle U_z \rangle _A =  \unit{0.5}{\meter \per \second}$  and study the influence of the active grid on the velocity statistics. 
By varying the grid speed factor (GSF) from 0 to 1, it is easily observed that the active grid has a strong influence on the velocity fluctuations. 
For a grid speed factor (GSF) equal to zero, the active grid is switched off, and the rods are placed in such a way that the flaps are open, forming a passive grid instead. 
In this case we observe the weakest velocity fluctuations. 
As the GSF increases the fluctuations increase up to 3 times, and for each of the cases the velocity fluctuations remain nearly constant along the width in the bulk.
In Figure \ref{fig:uz_pdf} we show the probability density function of the streamwise velocity obtained from LDA measurements for different GSF at the centre of the setup ($x/W = 0.5$, $y/D = 0.5$, $z/L = 0.5$) at mean flow rate of $\unit{0.5}{\meter \per \second}$. 
We observe that for all the settings of the active grid the distribution of the vertical velocity nearly follows a normal distribution, with a skewness of around 0.2 and a kurtosis of around 3.25 for all cases.

Lastly, we examine the homogeneity in the wall-normal direction by measuring the vertical velocity along the depth of the setup at  $z/L = 0.5$,  $\langle U_z \rangle _A = \unit{0.5}{\meter \per \second}$ and $GSF = 0.5$ using LDA.
Results presented in Figure \ref{fig:uz_width} demonstrate that the velocity is nearly constant in the wall-normal direction as well, with variations of only around $\pm 2\%$ from $y/D = 0.2$ to $y/D = 0.8$.

%----------------------------------------------------
\subsection{Local temperature measurements}
%----------------------------------------------------
\begin{figure}
	\centering
	\includegraphics[scale=0.4]{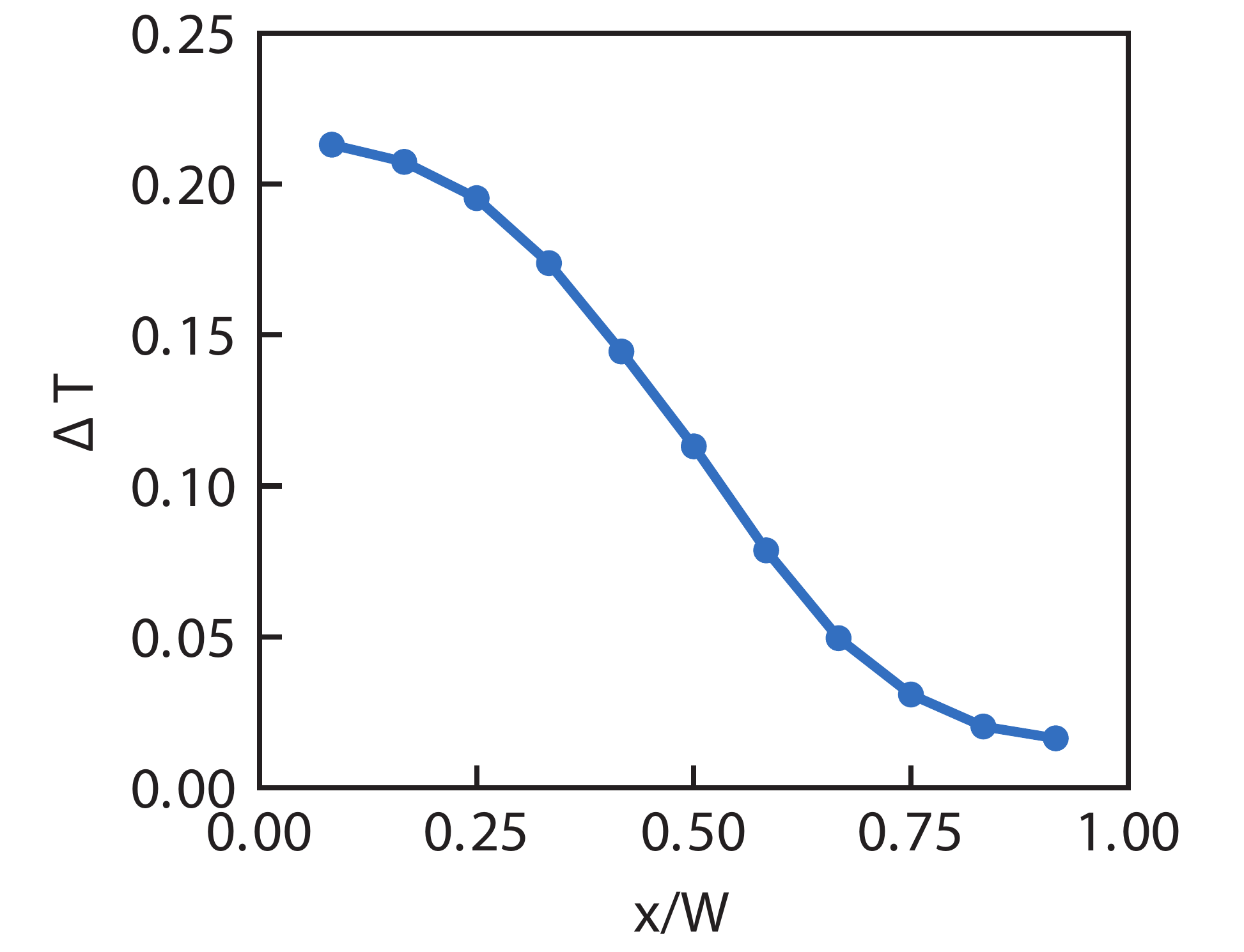}
	\caption{\label{fig:T_prof} Temperature profile at half height in the case of heating of the left half $x/W = 0$ to $x/W = 0.5$ of the setup. Here $\Delta T = T_{inlet} - T$, where $T_{inlet}$ is the inlet temperature measured before the heaters, and $T$ is the temperature measured by a thermistor in the measurement section.}
\end{figure}

\begin{figure}
	\centering
	\includegraphics[scale=0.4]{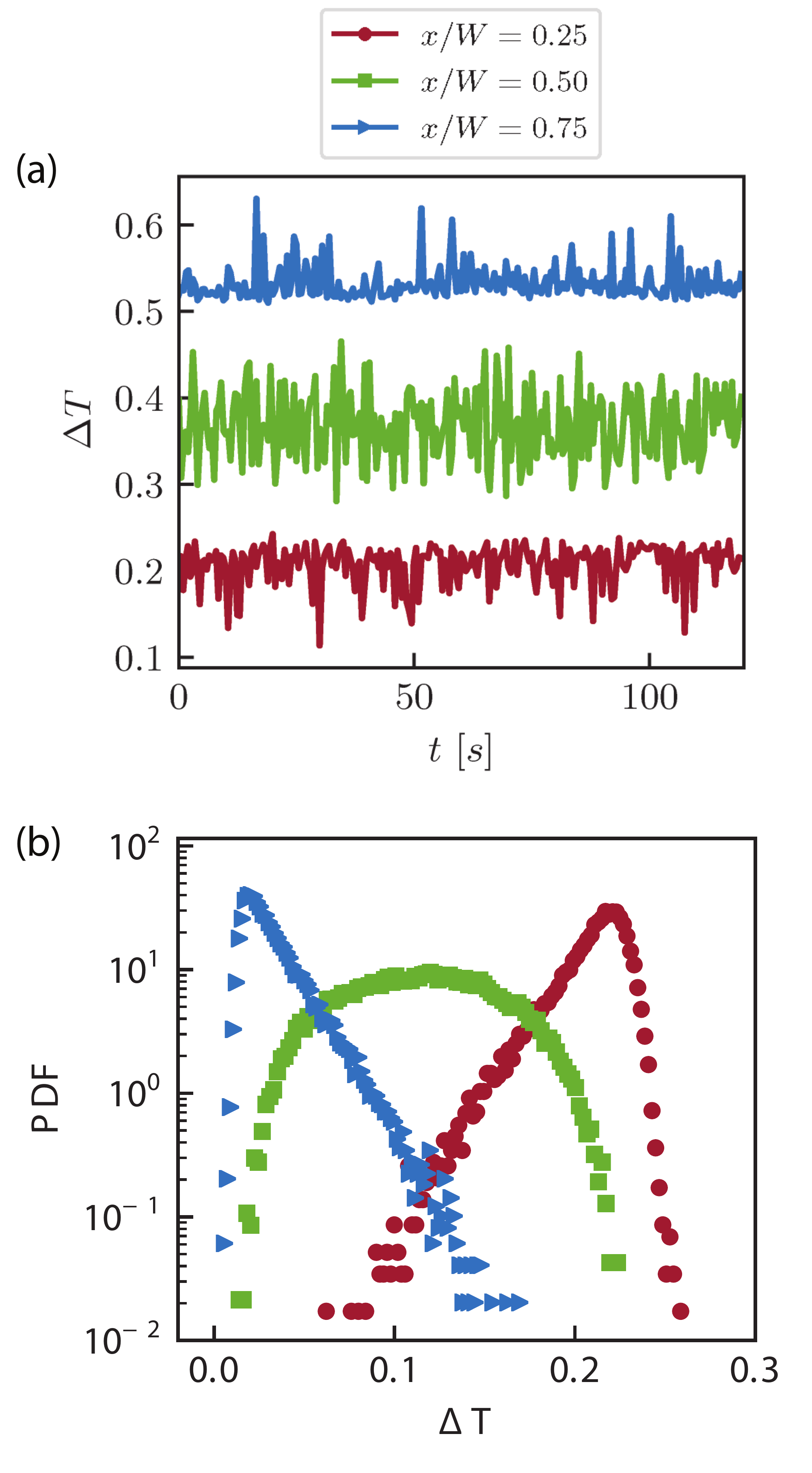}
	\caption{\label{fig:T_s} (a) Temperature series measured at different locations: $x/W= 0.25$ above the heated area, $x/W= 0.50$ in the centre of the setup and $x/W= 0.75$ above the non-heated area. Note that for the signals at $x/W= 0.50$ and $x/W= 0.75$ offsets of 0.25 and 0.50, respectively, are applied, for clarity of reading the figure. (b) Probability density function of the temperature at different locations of the setup: $x/W= 0.25$ above the heated area, $x/W= 0.50$ in the centre of the setup and $x/W= 0.75$ above the non-heated area.}
\end{figure}

Here we examine the influence of heating of one half of the heating section on the temperature profile in the measurement section of dimensions width $\times$ depth $= 0.3 \times \unit{0.04}{\meter ^2}$, with the mean liquid flow of $\unit{0.5}{\meter \per \second}$ and the grid speed factor of 0.5.
In this specific configuration six heaters (out of twelve total) which are placed in the left half ($x/W \leq 0.5$) of the heating section have been supplied by a constant power of $\unit{870}{\watt}$ each.
The thermistor for the temperature measurements is inserted to the measurement section from the sidewall of the section and traversed over the width at half height.
The measurements are logged at $\unit{2}{\hertz}$ by a digital multimeter over 2 hours after statistically steady state is achieved.
Figure \ref{fig:T_prof} shows temperature profile obtained in such a way, where heating was introduced in the heating section $x/W \leq 0.5$. 
Here $\Delta T = T_{inlet} - T$ is the difference between temperature measured before the heaters $T_{inlet}$ and the temperature measured in the bulk in the measurement section $T$.
We observe decrease in $\Delta T$ from $x/W = 0$ to $x/W = 1$ as a consequence of heating of only one half of the section.

After passing the heated section, warm and cold volume of liquid mix, resulting in different temperature signals in the measurement section at $x/W = 0.25$, $x/W = 0.5$ and  $x/W = 0.75$ (see Figure \ref{fig:T_s}a)).
This is also reflected on the probability density function shown in figure \ref{fig:T_s}b). 
Namely, negative skewness is observed for the signal at $x/W = 0.25$ due to non-heated liquid penetrating to the opposite side of the section.
In the center ($x/W = 0.5$), where the mixing between cold and warm part of the bulk is the most intense, the signal is not skewed (skewness $=0$), i.e. positive and negative fluctuations are equally represented. 
Again at $x/W = 0.75$ warm liquid parcels penetrate the cold part of the bulk, yielding a positively skewed PDF.
%----------------------------------------------------
\subsection{Dispersion of a passive scalar in a turbulent bubbly flow}
%----------------------------------------------------

The TMHT offers the possibility of studying the mixing of a passive scalar in turbulent bubbly flow. 
The primary advantage of this setup over, for example, the Twente Water Tunnel \cite{poorte1998thesis} is that in this setup we can easily achieve gas volume fractions higher than $1\%$. 
For the purpose of studying mixing of a low-diffusive dye induced by a swarm of high Reynolds number bubbles rising within a turbulent flow we inject a fluorescent dye (Fluorescein sodium) in the tunnel with the measurement section $\#$2 (width $\times$ depth $= \unit{0.3 \times 0.06}{\meter^2}$). 
The dye injector (diameter \unit{2}{\milli \meter}) is placed in the middle of the cross-section \unit{20}{\centi\meter} away from the bottom of the measurement section.
The injection is performed over 60 seconds at a flow rate matching the mean liquid velocity in the tunnel (\unit{\sim 30}{\centi\meter\per \second}).
The images of the flow are taken 10 second after the start of the injection, using two synchronized and vertically aligned cameras (Imager sCMOS, Lavision) with \unit{105}{\milli \meter} macro lenses and optical band-pass filters (450--650 nm). 

\begin{figure*}
	\centering
	\includegraphics[scale=0.4]{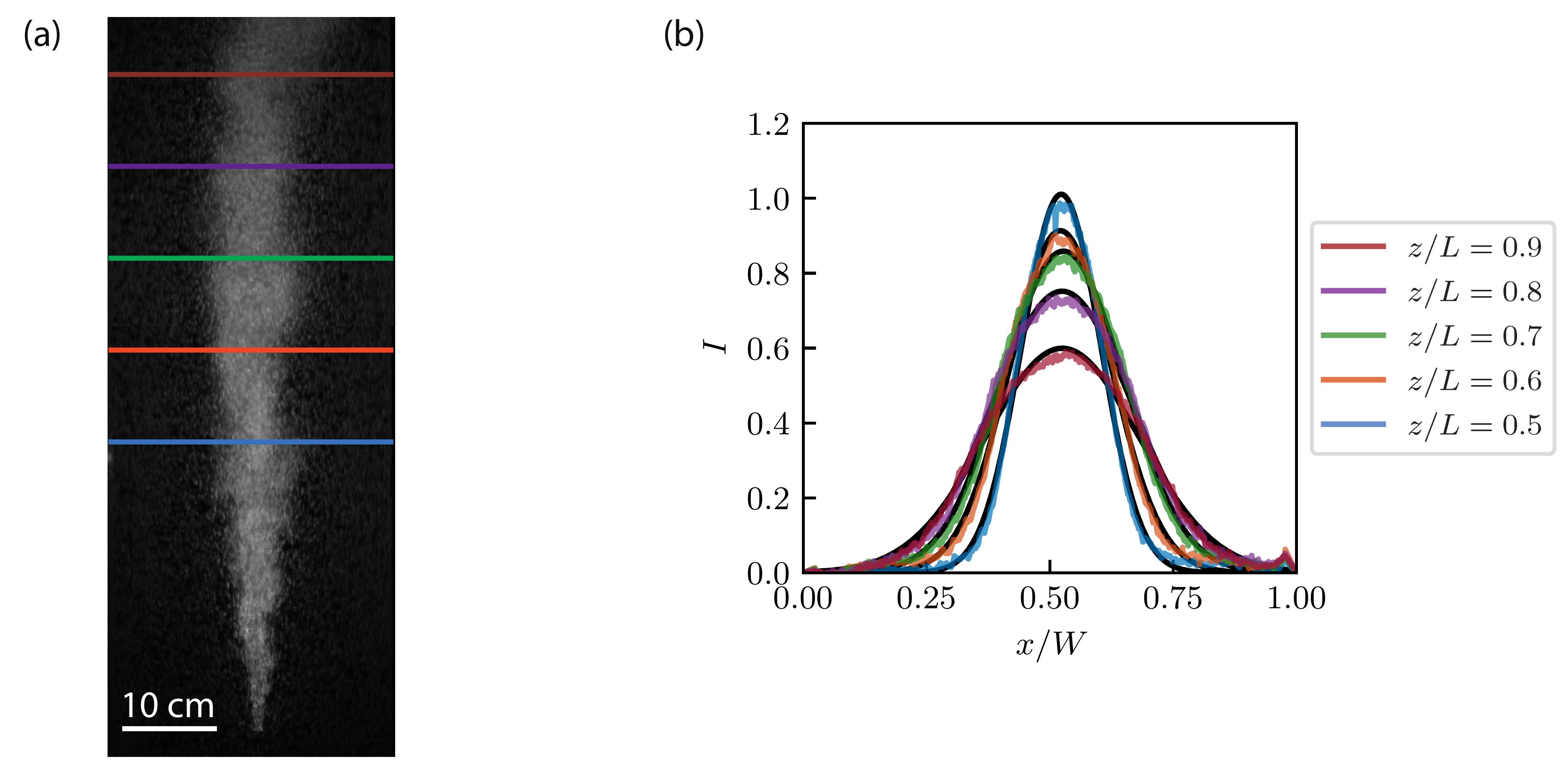}%ection_2.pdf}
	\caption{\label{fig:dye} Dispersion of a fluorescent dye in a bubbly turbulent flow with $\langle U_z \rangle_A =\unit{0.30}{\meter \per \second}$. (a) Instantaneous snapshot of the flow taken at an arbitrary time showing the spatial distribution of the fluoresced light. (b) Time averaged light intensity profiles across the width of the setup at different heights (in colour) and corresponding Gaussian fits. }
\end{figure*}

In figure \ref{fig:dye}a, we show an instantaneous snapshot of the fluorescent dye being advected and diffused in the presence of a bubbly turbulent flow. 
Using a time--series of such snapshots, we measure the mean intensity profiles of the fluorescent light $I$ across the width of the setup at different heights; profiles for a few selected heights are shown in figure \ref{fig:dye}b. 
The profiles are symmetric due to the homogeneity of the flow in the horizontal direction. 
Additionally, we find that the profiles are nearly normally distributed, suggesting that the spreading of the dye in the horizontal dimension is dominated by diffusion. 
The objective of future studies is to obtain the horizontal diffusion coefficient by measuring concentration of the dye by means of quantitative laser induced fluorescence. 
For further details on this experimental technique we refer to Alm{\'e}ras \emph{et al.} \cite{almeras2015mixing}.

%----------------------------------------------------
\subsection{Salt}
%----------------------------------------------------

\begin{figure*}
    \centering
    \includegraphics[scale=2.1]{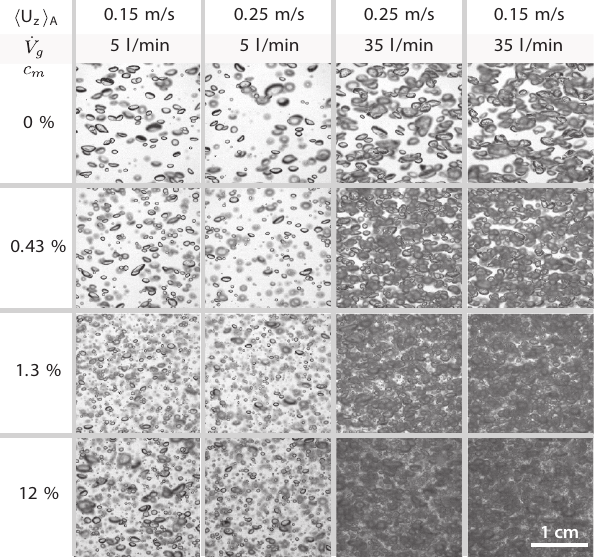}
    \caption{\label{fig:salt} Snapshots of the flow in the TMHT with the addition of increasing salt concentration (from top to bottom), for various water flows and gas flow rates. $\langle U_z \rangle _A$ is the mean liquid streamwise velocity, $\dot V_g$ is the gas flow rate, $c_m$ is the mass fraction of salt in the solution.}
\end{figure*}

As discussed in the section \ref{sec:intro}, the TMHT also allows the possibility to study the dynamics of bubbles injected in brine (salt solution). 
We perform preliminary tests by varying the mean liquid flow rate, gas flow rate, and the concentration of salt (NaCl) in the brine solution with a constant grid speed factor $GSF = 0.4$, while using the measurement section $\#$2.
In figure \ref{fig:salt}, we show snapshots for the different cases. 
Images were taken using a Photron SA1 camera with 1000 frames per second.

We visually observe a big difference in the flow regimes as we increase the salt concentration of the solution from $c_m=0\%$ to $c_m=12\%$ (mass fraction). 
The effect is more pronounced when the gas flow rate is higher, which may be due to presence of more smaller bubbles caused by the increase in surface tension by the addition of salt. 
This also leads to a reduction in the deformability of the bubbles and consequently to a change in the dynamics of the system.

 \begin{figure*}
 	\centering
 	\includegraphics[scale=0.35]{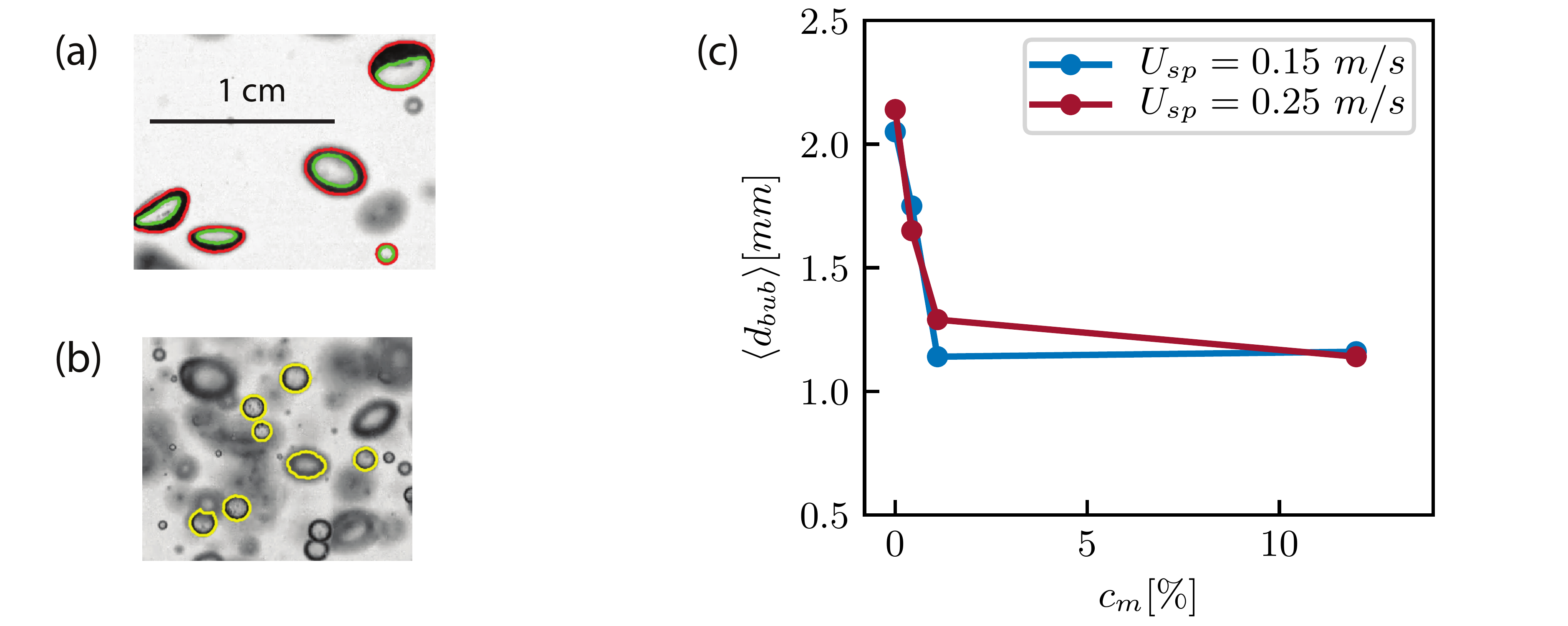}
 	\caption{\label{fig:salt_res} (a) Zoom of a snapshot for the case of mean liquid velocity $\langle U_z\rangle_A = 0.15 $~m/s, gas flow rate $ \dot{V} = 5 $~l/min and $c_m = 0 \%$, showing bubbles identified as pairs of objects; (b) Zoom of a snapshot for the case of  $\langle U_z\rangle_A = 0.15 $~m/s, gas flow rate $ \dot{V} = 5 $~l/min and $c_m = 12 \%$, showing bubbles identified as round objects; (c) Mean bubble diameter for $\langle U_z\rangle_A = 0.15$~m/s and $\langle U_z\rangle_A = 0.25$~m/s for varying salt concentrations.}
 \end{figure*}
 
By means of image processing, we measure the bubble diameter for the cases of mean liquid velocity 0.15~m/s and 0.25~m/s and gas flow rate of 5 l/min for varying salt concentrations. 
In figures \ref{fig:salt_res}a and \ref{fig:salt_res}b, we show sample images from two different simple image processing algorithms for edge detection implemented using Matlab.
For cases with low concentration of salt ($c_m \leq 1.3\%$) we identify contours of objects as well as boundaries of holes inside these objects (see Figure \ref{fig:salt_res}a).
The criterium used to identify bubbles is that the area of the outermost object is 1.3 times greater than the area of the object completely enclosed by it.
For the case of high salt concentration $c_m = 12\%$, for which more spherical bubbles are present, bubbles are selected by identifying round objects under the condition that the bubbles are in focus (see Figure \ref{fig:salt_res}b). 
Each method detects at least 1300 bubbles for each of the examined cases. 
Average bubble diameters obtained this way are shown in figure \ref{fig:salt_res}c. 
We find that a small increase of the concentration of salt from $c_m = 0 \%$ to $c_m = 1.3\%$ significantly decreases the average bubble diameter. 
Further increasing the salt concentration from  $c_m = 1.3\%$ to $c_m = 12\%$ seems to have hardly any influence on the average bubble diameter.
For each examined salt concentration, an increase in the mean liquid flow rate resulted in almost no  change in the average bubble diameter. 
Future studies in this setup will be dedicated to studying the effect of salt on the dynamics of the system and the coupling of heat transport with the salt concentration.

%%%%%%%%%%%%%%%%%%%%%%%%%%%%%%%%%%%%%%%%%%%%%%%
\section{\label{sec:outlook}Summary and outlook}%
%%%%%%%%%%%%%%%%%%%%%%%%%%%%%%%%%%%%%%%%%%%%%%%
A new experimental facility the Twente Mass and Heat Transfer Water Tunnel  (TMHT) has been built. 
This facility has global temperature control, bubble injection and local heat/mass injection and offers the possibility to study heat and mass transfer in turbulent multiphase flow.
The tunnel is made of high-grade stainless steel permitting the use of salt solutions in excess of 15$\%$ mass fraction, besides water. 
The total tunnel volume is \unit{300}{\liter}. 
Three interchangeable measurement sections of $1$ m height but of different cross sections ($0.3 \times 0.04$ m$^2$, $0.3 \times 0.06$ m$^2$, $0.3 \times 0.08$ m$^2$) span a Reynolds-number range from  $1.5 \times 10^4$ to $3 \times 10^5$ in the case of water at room temperature.
The glass vertical measurement sections allow for optical access to the flow, enabling techniques such as laser Doppler anemometry, particle image velocimetry, particle tracking velocimetry, and laser-induced fluorescent imaging. 
Thermistors mounted on a built-in traverse provide local temperature information at a few milli-Kelvin accuracy. 
Combined with simultaneous local velocity measurements, the local heat flux in single phase and two phase turbulent flow can be studied.

We demonstrated long term and short term stability of the power of the heaters and the cooler, of the inlet and outlet temperature of the tunnel, and the mean flow rate. 
PIV and LDA measurements were performed for a variety of flow conditions to show that homogeneity of the flow is satisfying. 
Preliminary temperature measurements in single phase flow and measurements with salt and dye injection in two phase flows were performed as well.

\FloatBarrier

\begin{acknowledgments}
This work is part of the Industrial Partnership Programme i36 Dense Bubbly Flows that is carried out under an agreement between  AkzoNobel (Nouryon since October 2018), DSM Innovation Center B.V., SABIC Global Technologies B.V., Shell Global Solutions B.V., TATA Steel Nederland Technology B.V. and the Netherlands Organisation for Scientific Research (NWO). 
Chao Sun acknowledges the financial support from Natural Science Foundation of China under Grant No. 11672156.
This work was also supported by The Netherlands Center for Multiscale Catalytic Energy Conversion (MCEC), an NWO Gravitation Programme funded by the Ministry of Education, Culture and Science of the government of The Netherlands. 
Authors thank Bert Vreman for his contributions to the measurements with salt, Varghese Mathai for fruitful discussions and useful insights, Rodrigo Ezeta for helping set up the PIV measurements, and Bas Dijkhuis and Hein-Dirk Smit for participating in the velocity and temperature measurements.

\end{acknowledgments}

%\bibliography{RSIbib}

%

\end{document}